\pdfoutput=1

\documentclass[11pt,aps,prd,amssymb,amsmath,tightenlines]{revtex4}
\usepackage{graphicx,epsfig}

\def\half{{\textstyle{\frac{1}{2}}}}
\def\quarter{{\textstyle{\frac{1}{4}}}}
\def\eighth{{\textstyle{\frac{1}{8}}}}
\def\threequarters{{\textstyle{\frac{3}{4}}}}
\def\threehalf{{\textstyle{\frac{3}{2}}}}

\def\cPT{\mathcal{PT}}

\begin{document}

\title{Fourth Painlev\'e Equation and $\cPT$-Symmetric Hamiltonians}

\author{Carl M. Bender}\email{cmb@wustl.edu}
\affiliation{Department of Physics, Washington University, St. Louis, MO
63130, USA}
\author{Javad Komijani}
\email{jkomijani@ut.ac.ir}
\affiliation{Department of Physics, University of Tehran, Tehran 1439955961,
Iran}

\date{\today}

\begin{abstract}
This paper is an addendum to earlier papers \cite{R1,R2} in which it was shown
that the unstable separatrix solutions for Painlev\'e I and II are determined by
$\cPT$-symmetric Hamiltonians. In this paper unstable separatrix solutions of
the fourth Painlev\'e transcendent are studied numerically and analytically. For
a fixed initial value, say $y(0)=1$, a discrete set of initial slopes $y'(0)=
b_n$ give rise to separatrix solutions. Similarly, for a fixed initial slope,
say $y'(0)=0$, a discrete set of initial values $y(0)=c_n$ give rise to
separatrix solutions. For Painlev\'e IV the large-$n$ asymptotic behavior of
$b_n$ is $b_n\sim B_{\rm IV}n^{3/4}$ and that of $c_n$ is $c_n\sim C_{\rm IV}
n^{1/2}$. The constants $B_{\rm IV}$ and $C_{\rm IV}$ are determined both
numerically and analytically. The analytical values of these constants are found
by reducing the nonlinear Painlev\'e IV equation to the linear eigenvalue
equation for the sextic $\cPT$-symmetric Hamiltonian $H=\half p^2+\eighth x^6$.
\end{abstract}

\maketitle

\section{Introduction}\label{s1}
The six Painlev\'e transcendents satisfy nonlinear second-order differential
equations having the property that their movable (spontaneous) singularities are
poles (and not branch points or essential singularities). Many papers have been
published on these equations (see, for example, References~1-8 in \cite{R2} for
background information and References~9-16 in \cite{R2} for applications in
mathematical physics). This paper considers the fourth Painlev\'e transcendent,
referred to here as P-IV. There are many recent studies of this equation; see,
for example, Refs.~\cite{R3,R4,R5,R6}.

The initial-value problem for the P-IV differential equation examined here is
\begin{equation}
y(t)y''(t)=\half[y'(t)]^2+2t^2[y(t)]^2+4t[y(t)]^3+\threehalf[y(t)]^4,
\quad y(0)=c,~y'(0)=b.
\label{e1}
\end{equation}
(For simplicity we have set the two arbitrary constants in P-IV, one of which is
a trivial additive constant, to 0.) There have been many asymptotic studies of
the Painlev\'e transcendents, but here we present a simple numerical and
asymptotic analysis that has not appeared in the literature. This analysis
concerns the initial conditions that give rise to special unstable separatrix
solutions of P-IV. The asymptotic analysis here extends our earlier work on
nonlinear differential-equation eigenvalue problems in \cite{R2,R7,R8}.

The idea, originally proposed in Ref.~\cite{R6}, is that a nonlinear
differential equation may have a discrete set of {\it critical} initial
conditions that give rise to unstable separatrix solutions. These discrete
initial conditions can be thought of as eigenvalues and the separatrices
stemming from these initial conditions can be viewed as corresponding
eigenfunctions. The objective in Ref.~\cite{R6} is to find the large-$n$
(semiclassical) asymptotic behavior of t he $n$th eigenvalue. The analytical
approach is to reduce the nonlinear differential-equation problem to a linear
{\it problem} that could be solved to determine the asymptotic behavior of the
eigenvalues as $n\to\infty$.

A toy model used in Ref.~\cite{R6} to explore the properties of nonlinear
eigenvalue problems is the first-order differential-equation problem
\begin{equation}
y'(t)=\cos[\pi t\,y(t)],\quad y(0)=a.
\label{e2}
\end{equation}
The solutions to this initial-value problem exhibit $n$ maxima before vanishing
like $1/t$ as $t\to\infty$. As the initial condition $a$ increases past critical
values $a_n$, the number of maxima of $y(t)$ jumps from $n$ to $n+1$. At $a_n$
the solution $y(t)$ is an unstable separatrix: If $y(0)$ is slightly below
$a_n$, $y(x)$ merges with a bundle of stable solutions all having $n$ maxima and
when $y(0)$ is slightly above $a_n$, $y(x)$ merges with a bundle of stable
solutions all having $n+1$ maxima. We seek the asymptotic behavior of $a_n$ for
large $n$, which is the analog of a high-energy semiclassical approximation in
quantum mechanics. In Ref.~\cite{R6} it is shown that for large $n$ the
nonlinear differential equation (\ref{e2}) reduces to a {\it linear} difference
equation for a one-dimensional random walk. The difference equation is solved
exactly, and it is shown that
\begin{equation}
a_n\sim 2^{5/6}\sqrt{n}\quad(n\to\infty).
\label{e3}
\end{equation}
Kerr subsequently found an alternative solution to this asymptotics problem and
verified (\ref{e3}) \cite{R9}.

The nonlinear eigenvalue problem described above is similar in many respects to
the linear eigenvalue problem for the time-independent Schr\"odinger equation.
For a potential $V(x)$ that rises as $x\to\pm\infty$, the eigenfunctions $\psi(
x)$ of the Schr\"odinger eigenvalue problem
\begin{equation}
-\psi''(x)+V(x)\psi(x)=E\psi(x),\quad\psi(\pm\infty)=0,
\label{e4}
\end{equation}
are unstable with respect to small changes in the eigenvalue $E$; that is, if
$E$ is increased or decreased slightly, $\psi(x)$ abruptly violates the boundary
conditions (is not square integrable). Also, like the eigenfunctions (separatrix
curves) of (\ref{e2}), the $n$th eigenfunction $\psi_n(x)$ has $n$ oscillations
in the classically allowed region before decreasing monotonically to $0$ in the
classically forbidden region.

This paper considers two eigenvalue problems for P-IV. First, we find the
large-$n$ behavior of the positive eigenvalues $b_n$ for the initial condition
$y(0)=1,\,y'(0)=b_n$ and also the large-$n$ behavior of the negative eigenvalues
$c_n$ for the initial condition $y(0)=c_n,\,y'(0)=0$. We show that
\begin{equation}
b_n\sim B_{\rm IV}n^{3/4}\quad{\rm and}\quad c_n\sim C_{\rm IV}n^{1/2}.
\label{e5}
\end{equation}
In Sec.~\ref{s2} we compute the constants $B_{\rm IV}$ and $C_{\rm IV}$
numerically and in Sec.~\ref{s3} we find them analytically by reducing the
large-eigenvalue problem to the {\it linear} time-independent Schr\"odinger
equation for the sextic $\cPT$-symmetric Hamiltonian $H=\half p^2+x^6$.
Section~\ref{s4} gives brief concluding remarks.

\section{Numerical analysis of the fourth Painlev\'e transcendent} \label{s2}
There are three possible asymptotic behaviors of the solutions to the P-IV
equation as $t\to-\infty$; $y(t)$ can approach the straight lines $y=-2t$, $y=
-2t/3$, or $y=0$. An elementary asymptotic analysis shows that if $y(t)$
approaches $y=-2t/3$, the solution oscillates stably about this line with slowly
decreasing amplitude \cite{R10}. However, while $y=-2t$ and $y=0$ are possible
asymptotic behaviors, these behaviors are {\it unstable} and nearby solutions
veer away from them. Here we consider the eigenfunction solutions to P-IV that
approach $y=-2t$ as $t\to-\infty$. These separatrix solutions resemble
quantum-mechanical eigenfunctions because they have $n$ oscillations before
exhibiting this asymptotic behavior. Because the P-IV equation is nonlinear
these oscillations are unbounded; the $n$th eigenfunction passes through $[n/2]$
{\it simple poles} before smoothly approaching $y=-2t$ ($[n/2]$ is the greatest
integer in $\leq n/2$).

We consider two different eigenvalue problems for P-IV that are related to the
instability of the asymptotic behavior $y=-2t$: (i) We fix the initial value
$y(0)=1$ and seek the discrete values of the initial slopes $y'(0)=b$ that give
solutions approaching $-2t$, and (ii) we fix the initial slope $y'(0)=0$ and
seek the discrete initial values of $y(0)=c$ for which $y(t)$ approaches $-2t$.

\subsection{Initial-slope eigenvalues for Painlev\'e IV} \label{ss2a}
Let us examine the solutions to the initial-value problem for P-IV in (\ref{e1})
for $t<0$. As in Ref.~\cite{R2}, we find these solutions numerically by using
Runge-Kutta to integrate down the negative-real axis. When we approach a simple
pole, we integrate along a semicircle in the complex-$t$ plane around the pole
and continue integrating down the negative axis. We choose the initial value $y
(0)=1$ and allow the initial slope $y'(0)=b$ to have increasingly positive
values. (We only present results for positive initial slope; the P-IV equation
is symmetric under $t\to-t$ and also under $y\to-y$.) Numerical study shows that
the choice of $y(0)$ is not crucial if $y(0)\neq0$; for {\it any} $y(0)$ the
large-$n$ behavior of the initial-slope eigenvalues $b_n$ is the same.

\begin{figure}[h!]
\null\vspace{-9mm}
\begin{center}
\includegraphics[trim=21mm 50mm 18mm 50mm,clip=true,scale=0.45]{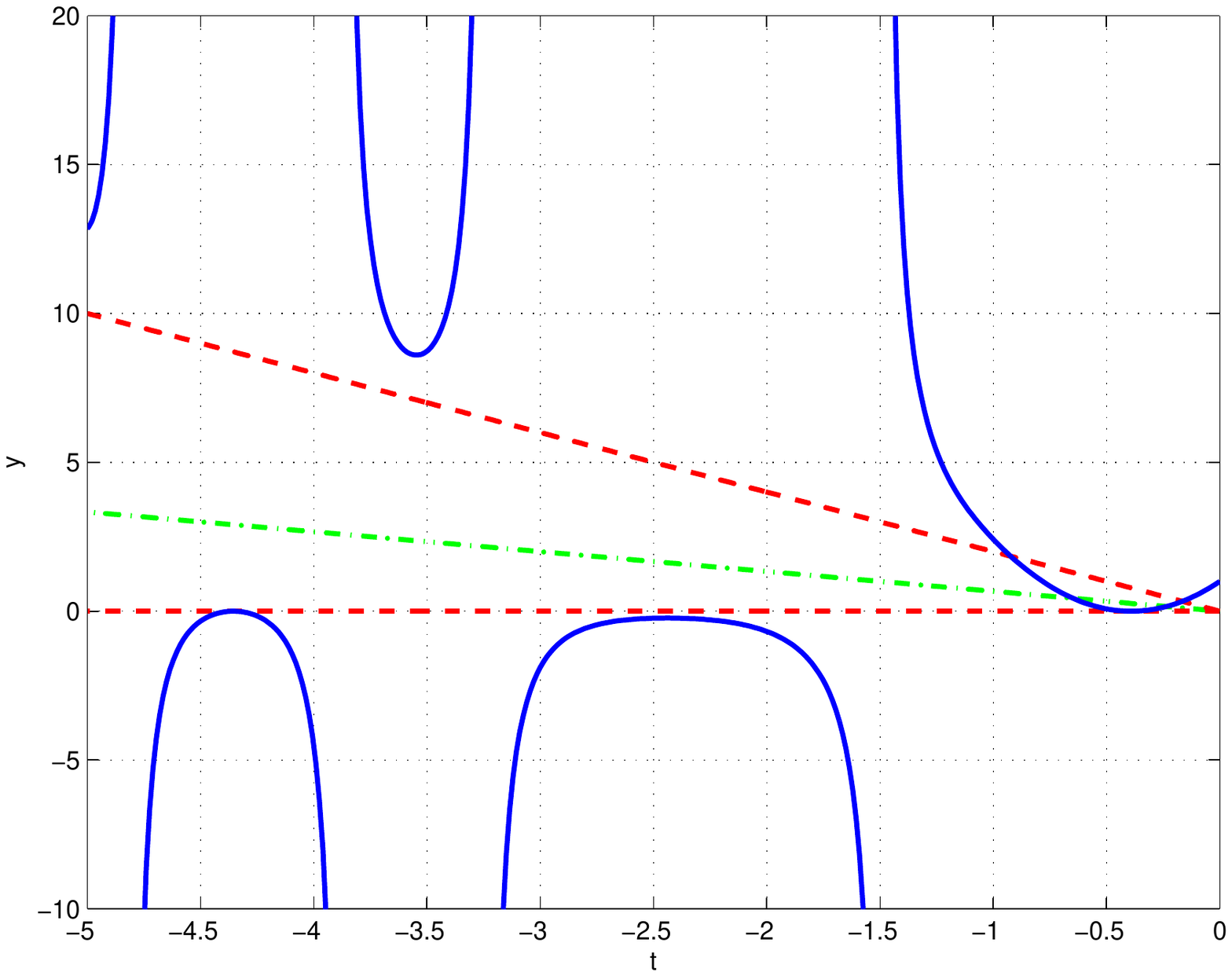}
% trim=left bottom right top
\includegraphics[trim=18mm 50mm 21mm 50mm,clip=true,scale=0.45]{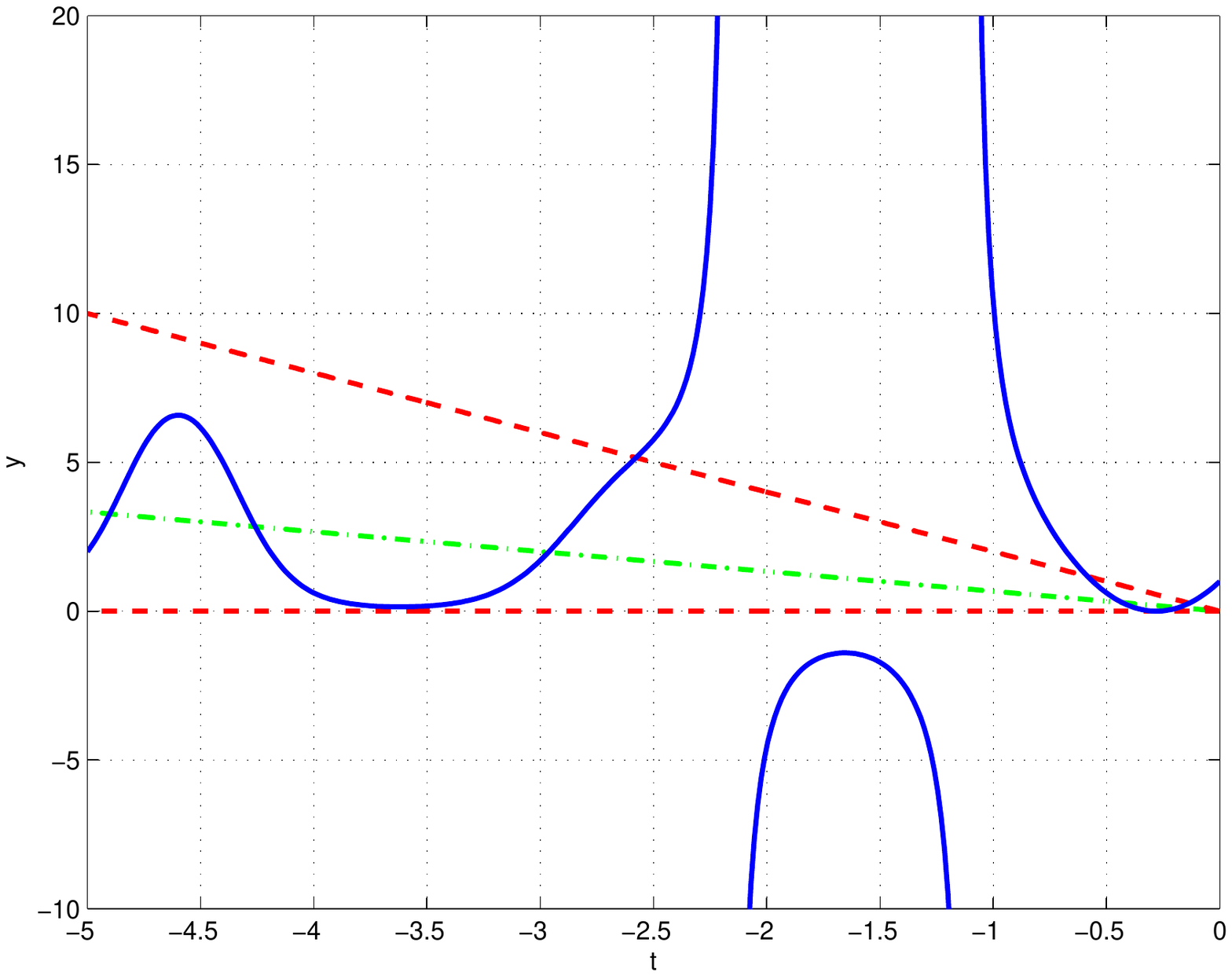}
\hspace{1.5cm}
\end{center}
\null\vspace{-21mm}
\caption{[Color online] Behavior of solutions $y(t)$ to the P-IV equation
(\ref{e1}) for initial conditions $y(0)=1$ and $b=y'(0)$. Left panel: $b=
5.18498704$, which lies between the eigenvalues $b_1=3.15837325$ and $b_2=
6.18498704$. Right panel: $b=7.18498704$, which lies between $b_2=6.18498704$
and $b_3=8.79172082$. The upper dashed line (red) is $y=-2t$, which is unstable
and the lower dashed line (stable) is $y=-2t/3$. In the left panel $y(t)$ has an
infinite sequence of simple poles but in the right panel the poles abruptly end
and solution then oscillates stably about $-2t/3$.}
\label{F1}
\end{figure}

\begin{figure}[h!]
\null\vspace{-9mm}
\begin{center}
\includegraphics[trim=21mm 50mm 18mm 50mm,clip=true,scale=0.45]{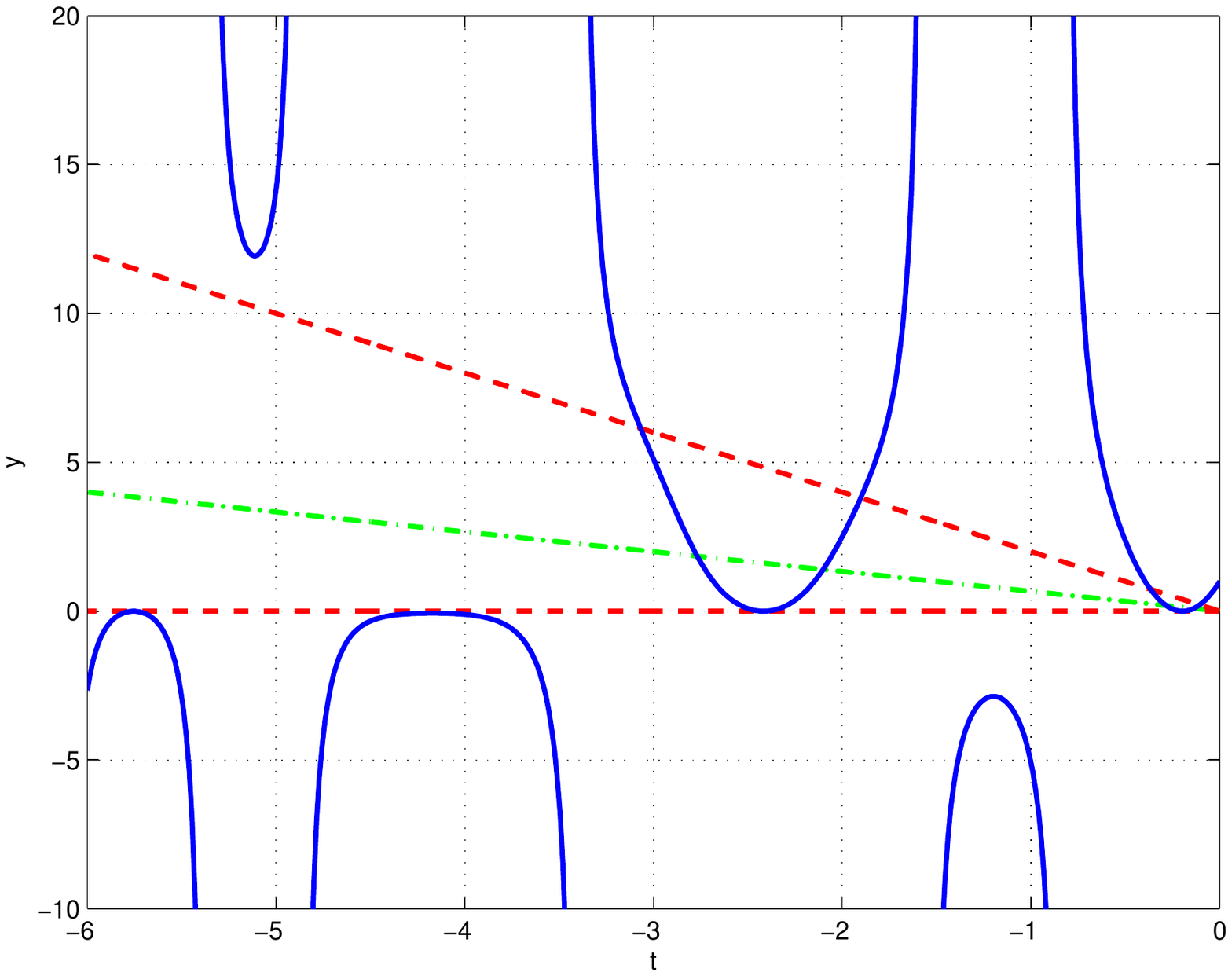}
% trim=left bottom right top
\includegraphics[trim=18mm 50mm 21mm 50mm,clip=true,scale=0.45]{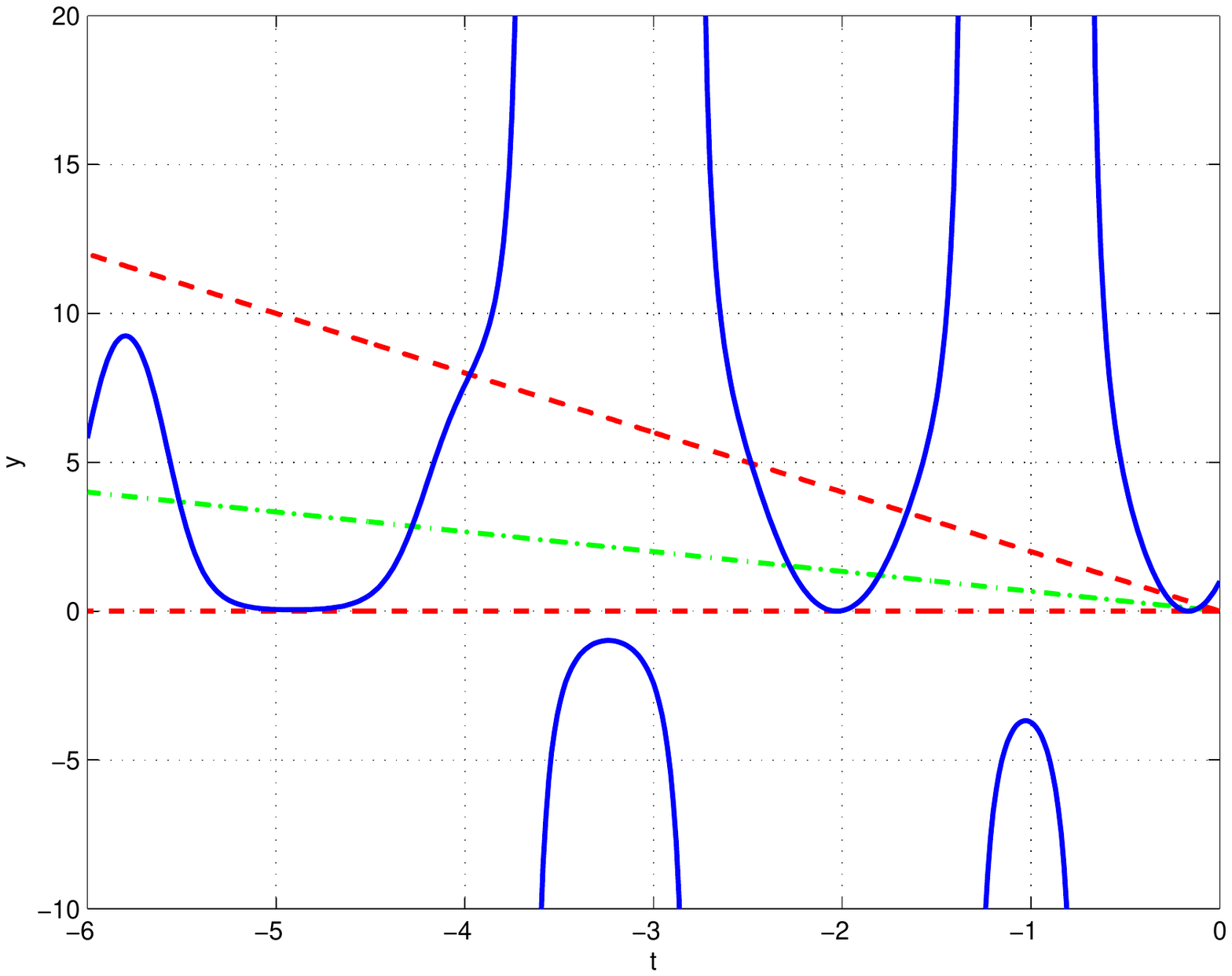}
\hspace{1.5cm}
\end{center}
\null\vspace{-21mm}
\caption{[Color online] Solutions to the P-IV equation (\ref{e1}) for $y(0)=0$
and $b=y'(0)$. Left panel: $b=10.1720921$, which lies between the eigenvalues
$b_3=8.79172082$ and $b_4=11.1720921$. Right panel: $b=12.1720921$, which lies
between the eigenvalues $b_4=11.1720921$ and $b_5=13.3990049$.}
\label{F2}
\end{figure}

Above the first eigenvalue $b_1=3.15837325$ there is a continuous interval of
$b$ for which $y(t)$ has an infinite sequence of simple poles (Fig.~\ref{F1},
left panel). When $b$ increases above the next eigenvalue $b_2=6.18498704$, the
character of the solutions changes abruptly and after $y(t)$ passes through a
finite number of simple poles it begins to oscillate stably about $-2t/3$
(Fig.~\ref{F1}, right panel). When $b$ exceeds the third eigenvalue $b_3=
8.79172082$, the solutions again pass through an infinite sequence of poles
(Fig.~\ref{F2}, left panel). When $b$ increases above $b_4=11.1720921$, the
solutions again oscillate stably about $-2t/3$ (Fig.~\ref{F2}, right panel).
Numerical study verifies that there is an infinite sequence of eigenvalues at
which the solutions to P-IV alternate between infinite sequences of simple poles
and stable oscillation about $-2t/3$.

\begin{figure}[h!]
\null\vspace{-9mm}
\begin{center}
\includegraphics[trim=21mm 50mm 18mm 50mm,clip=true,scale=0.45]{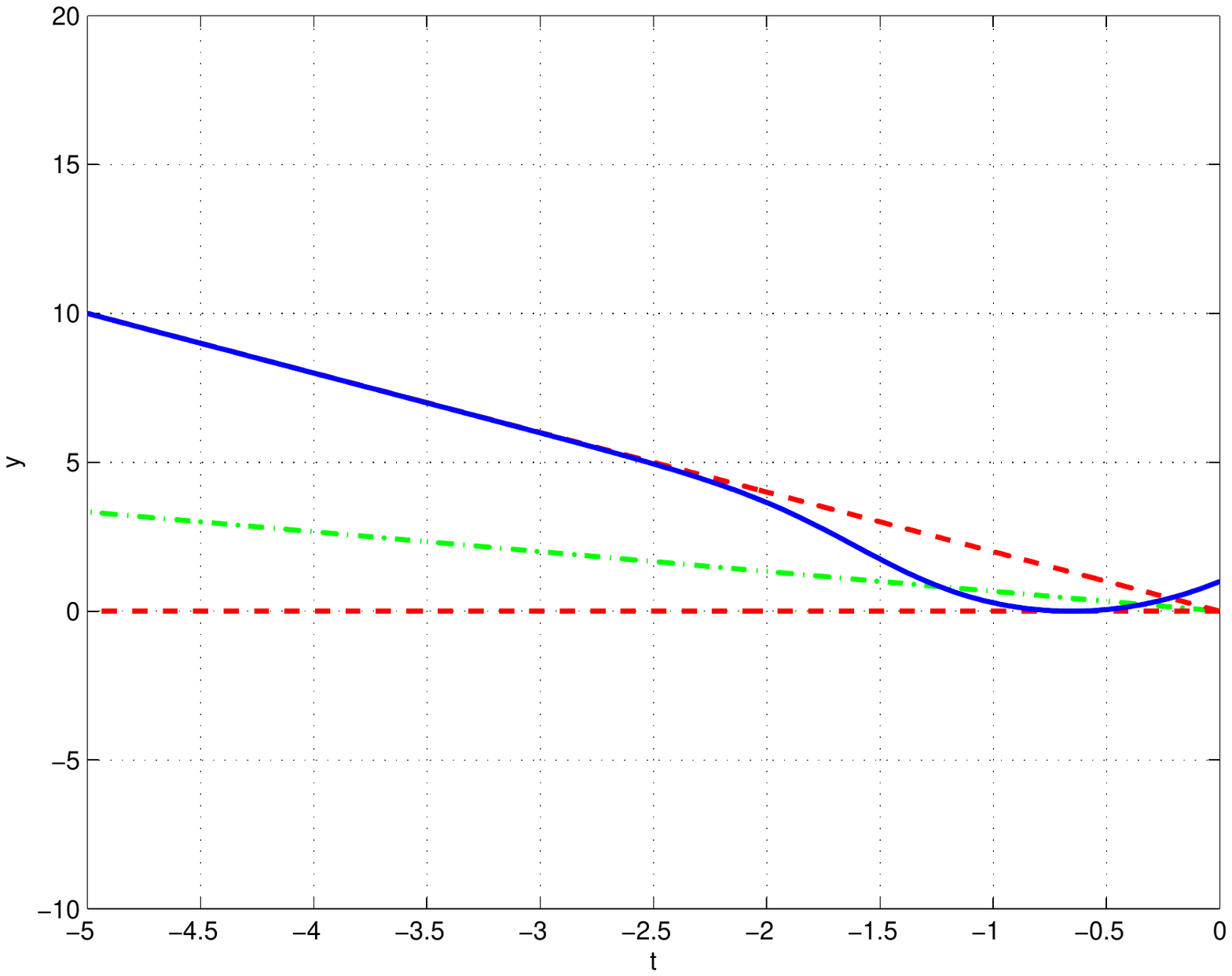}
% trim=left bottom right top
\includegraphics[trim=18mm 50mm 21mm 50mm,clip=true,scale=0.45]{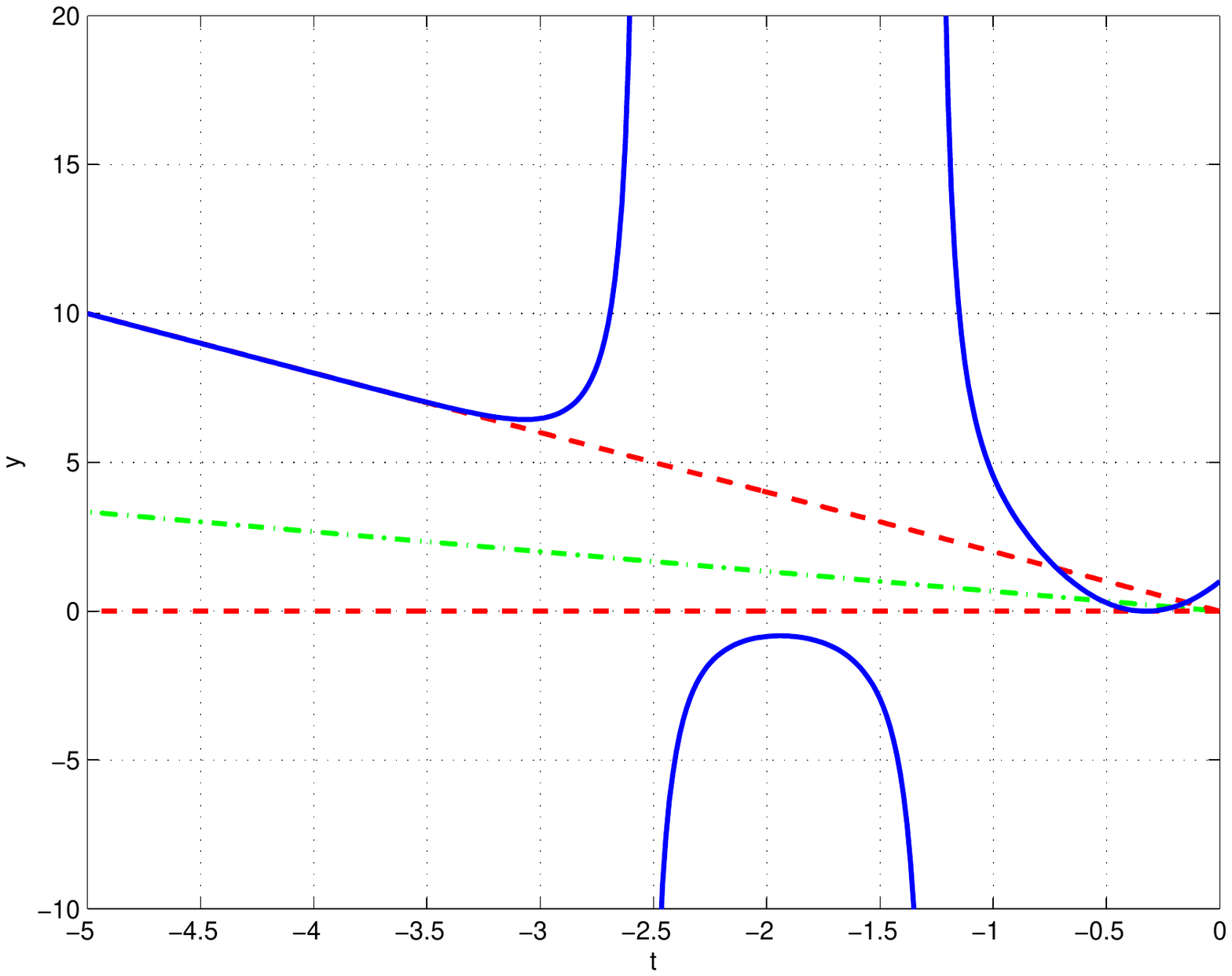}
\hspace{1.5cm}
\end{center}
\null\vspace{-21mm}
\caption{[Color online] First two separatrix (eigenfunction) solutions of P-IV
with initial condition $y(0)=0$. Left panel: $y'(0)=b_1=3.15837325$; right
panel: $y'(0)=b_2=6.18498704$. The dashed lines are $y=-2t$ and $y=-2t/3$.}
\label{F3}
\end{figure}

\begin{figure}[h!]
\null\vspace{-9mm}
\begin{center}
\includegraphics[trim=21mm 50mm 18mm 50mm,clip=true,scale=0.45]{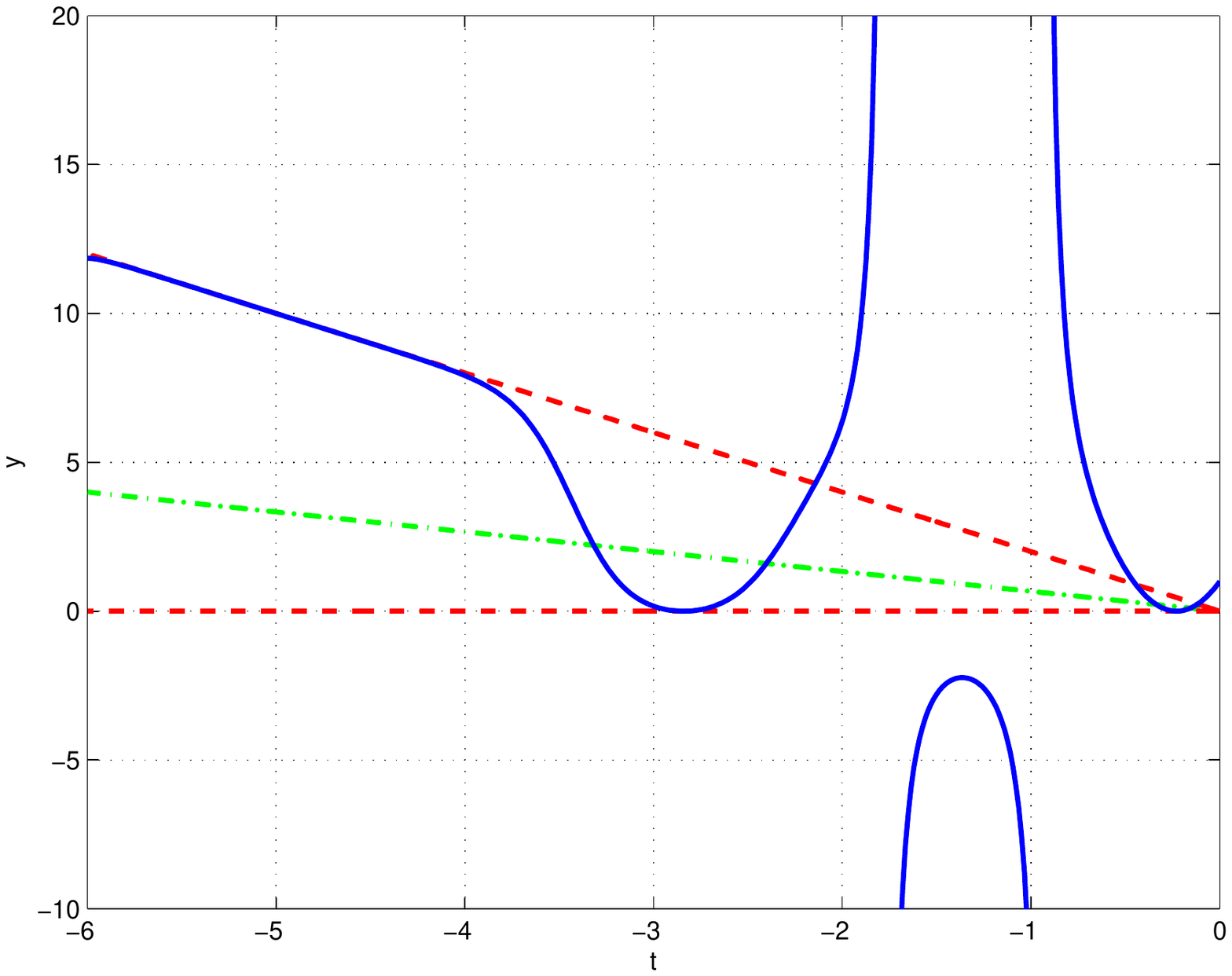}
% trim=left bottom right top
\includegraphics[trim=18mm 50mm 21mm 50mm,clip=true,scale=0.45]{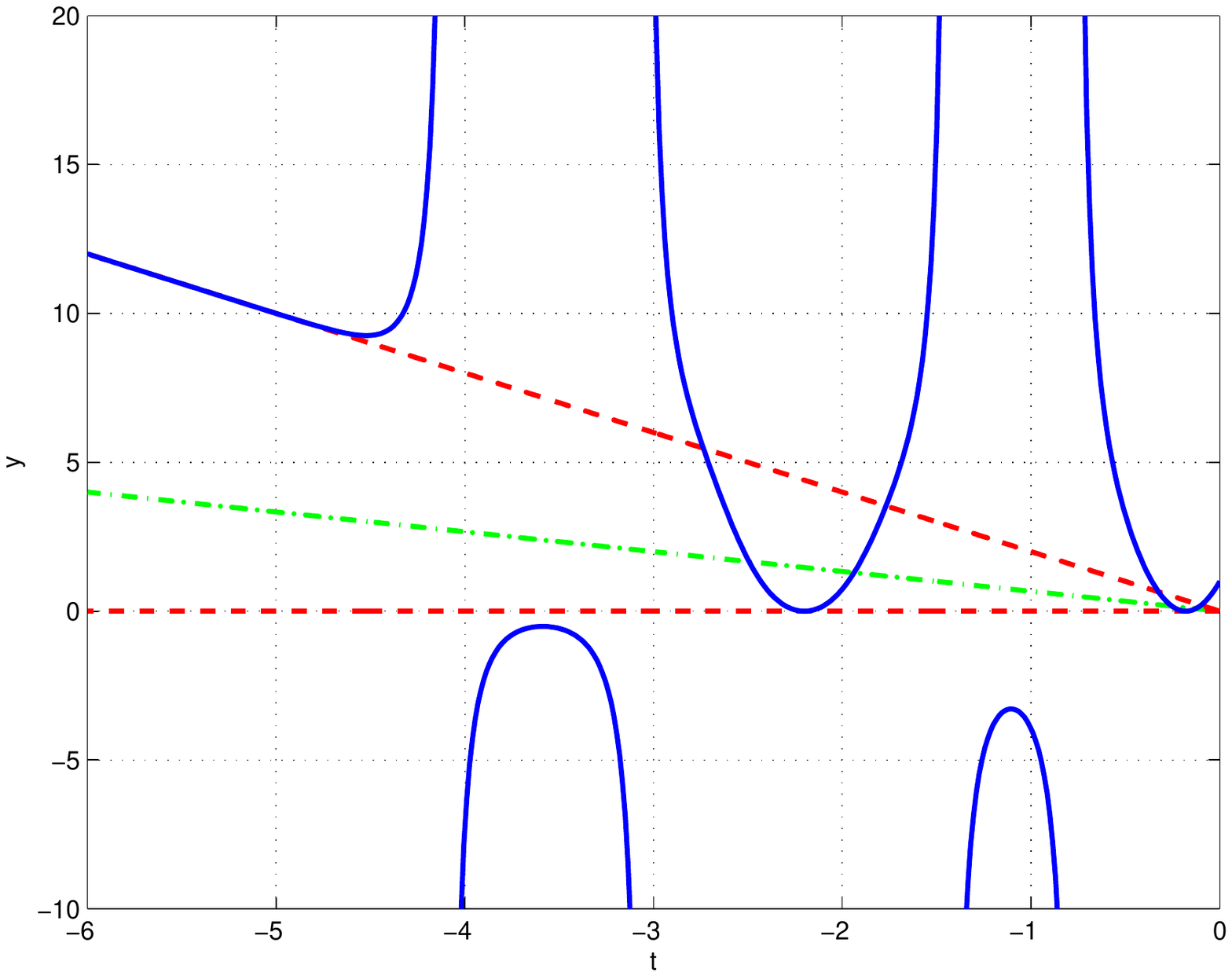}
\hspace{1.5cm}
\end{center}
\null\vspace{-21mm}
\caption{[Color online] Third and fourth eigenfunctions of P-IV with initial
condition $y(0)=0$. Left panel: $y'(0)=b_3=8.79172082$; right panel: $y'(0)=b_4
=11.1720921$.}
\label{F4}
\end{figure}

\begin{figure}[h!]
\null\vspace{-9mm}
\begin{center}
\includegraphics[trim=21mm 50mm 18mm 50mm,clip=true,scale=0.45]{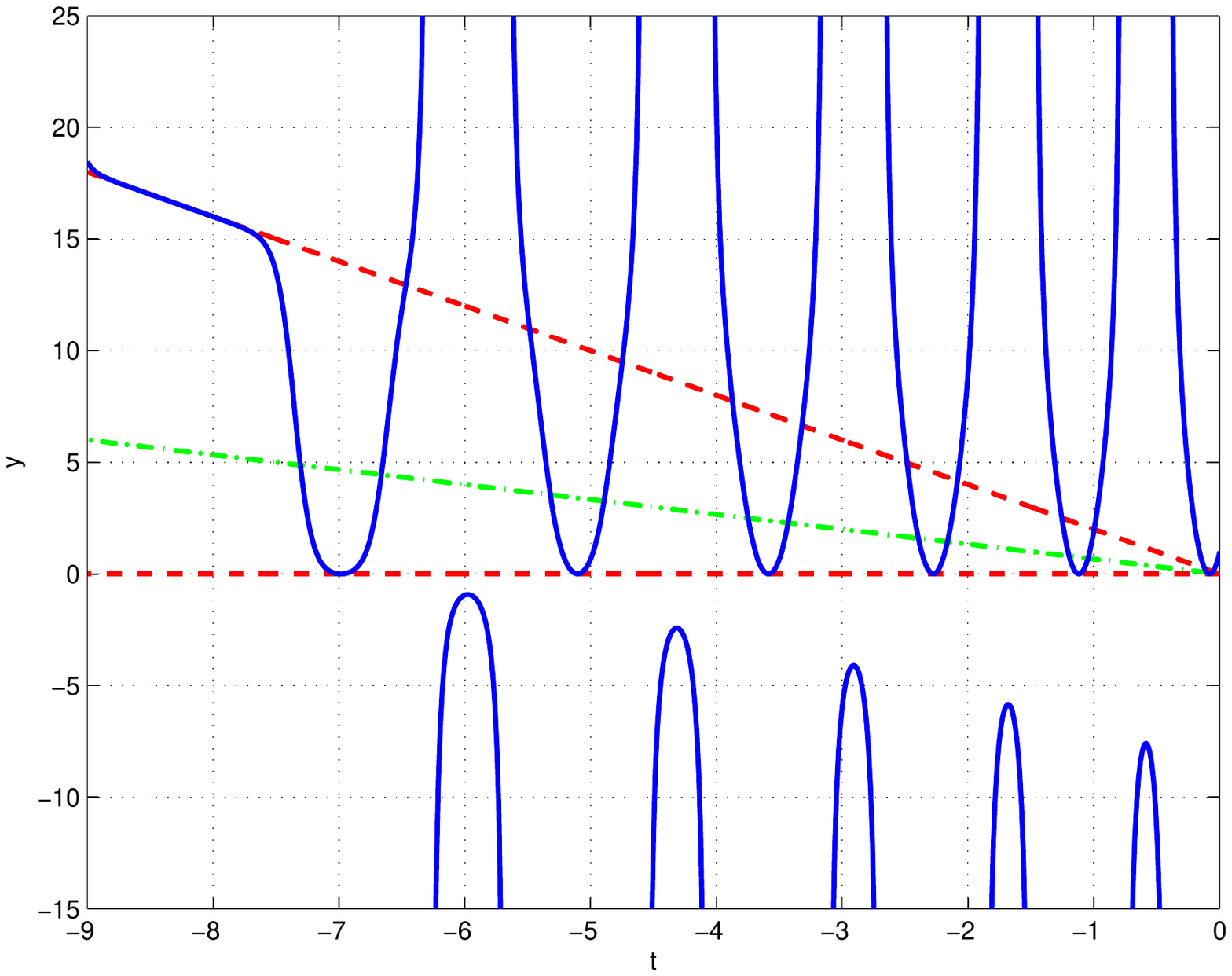}
% trim=left bottom right top
\includegraphics[trim=18mm 50mm 21mm 50mm,clip=true,scale=0.45]{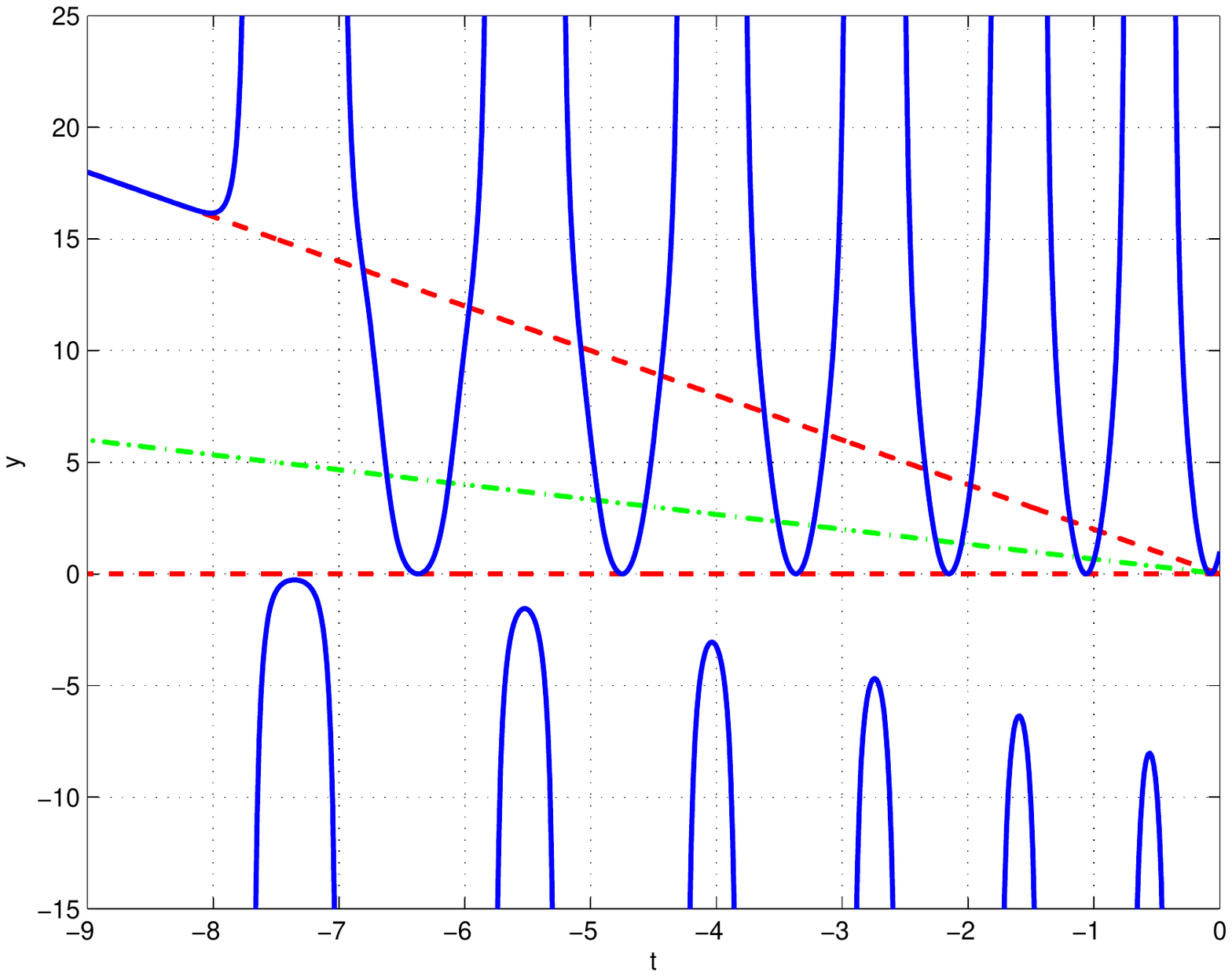}
\hspace{1.5cm}
\end{center}
\null\vspace{-21mm}
\caption{[Color online] Eleventh and twelth eigenfunctions of P-IV with initial
condition $y(0)=0$. Left panel: $y'(0)=b_{11}=24.9911479$; right panel: $y'(0)=
b_{12}=26.7370929$. As $n$ increases, the eigenfunctions pass through more and
more simple poles before exhibiting a turning-point transition and approaching
the limiting curve $-2t$ exponentially rapidly. This behavior is analogous to
that of the eigenfunctions of a time-independent Schr\"odinger equation for a
particle in a potential well; the higher-energy eigenfunctions exhibit more and
more oscillations in the classically allowed region before entering the
classically forbidden region, where they decay to zero.}
\label{F5}
\end{figure}

When $y'(0)$ is an eigenvalue the solutions exhibit a completely different and
unstable behavior from those in Figs.~\ref{F1} and \ref{F2}. These solutions
pass through a {\it finite} number of simple poles (like the oscillations of
quantum-mechanical eigenfunctions in a classically allowed region) and then have
a turning-point-like transition in which the poles cease and $y(t)$
exponentially approaches the line $-2t$. The solutions arising from the first
and second eigenvalues $b_1$ and $b_2$ are shown in Fig.~\ref{F3}, those
arising from $b_3$ and $b_4$ are shown in Fig.~\ref{F4}, and those arising from
$b_{11}$ and $b_{12}$ are shown in Fig.~\ref{F5}. The critical values $b_n$ are
analogous to eigenvalues because they generate {\it unstable} separatrix
solutions; if $y'(0)$ changes by a small amount above or below a critical value,
the character of the solutions changes abruptly and the solutions exhibit the
two possible generic behaviors shown in Figs.~\ref{F1} and \ref{F2}.

As in Ref.~\cite{R2} for P-I and P-II, we have performed a numerical asymptotic
study of the critical values $b_n$ for $n\gg1$ by using Richardson extrapolation
\cite{R10}. [In this paper we have taken $y(0)=1$ but we find that if $y(0)$ is
held fixed, the large-$n$ behavior of the initial slope $b_n$ is insensitive to
the choice of $y(0)$.] By applying fifth-order Richardson extrapolation to the
first twelve eigenvalues, we find the value of $B_{\rm IV}$ accurate to one part
in seven decimal places:
\begin{equation}
B_{\rm IV}=4.25684{\underbar 3}.
\label{e6}
\end{equation}

\subsection{Initial-value eigenvalues for Painlev\'e IV} \label{ss2b}
\begin{figure}[h!]
\null\vspace{-9mm}
\begin{center}
\includegraphics[trim=21mm 50mm 18mm
50mm,clip=true,scale=0.45]{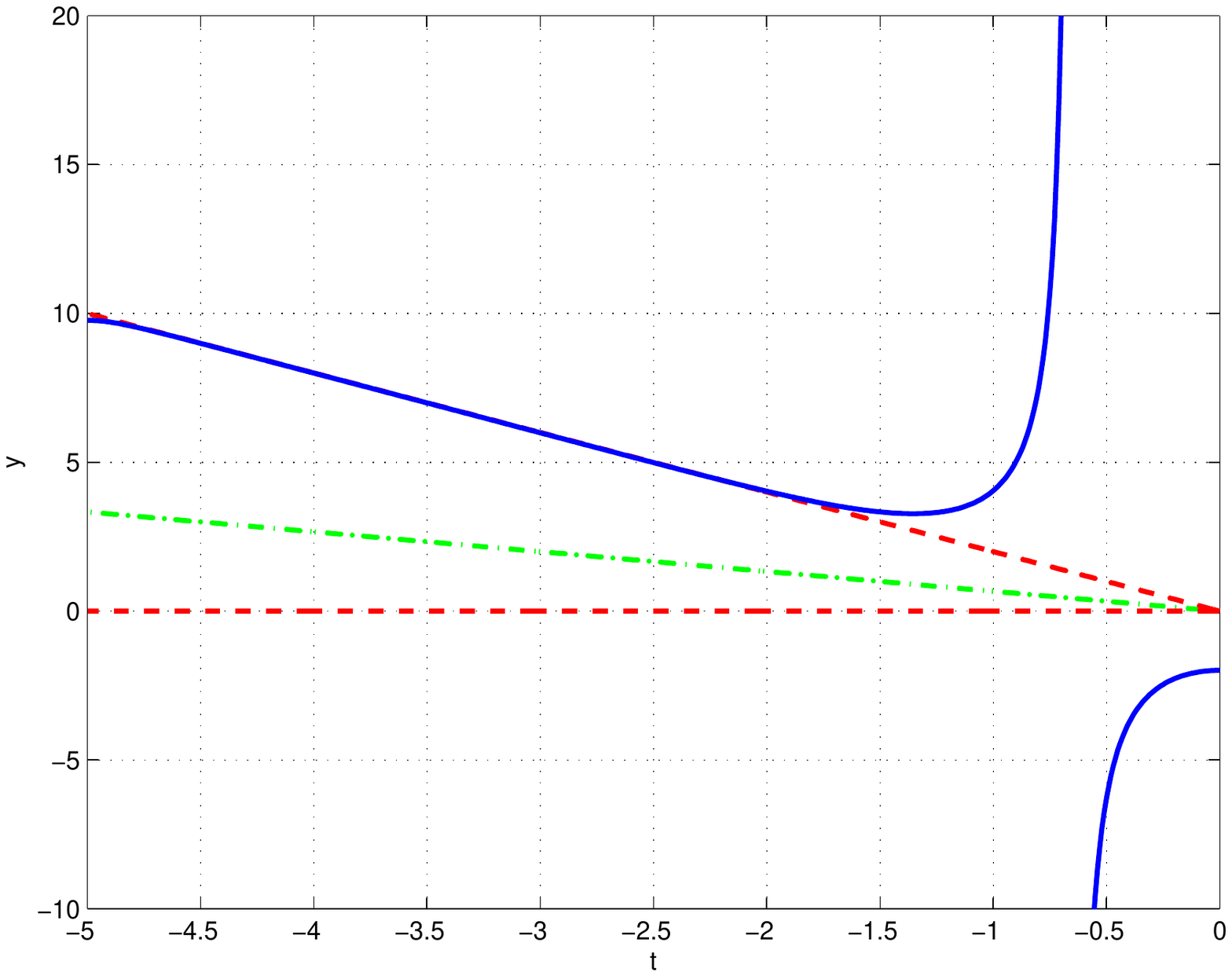}
% trim=left bottom right top
\includegraphics[trim=18mm 50mm 21mm
50mm,clip=true,scale=0.45]{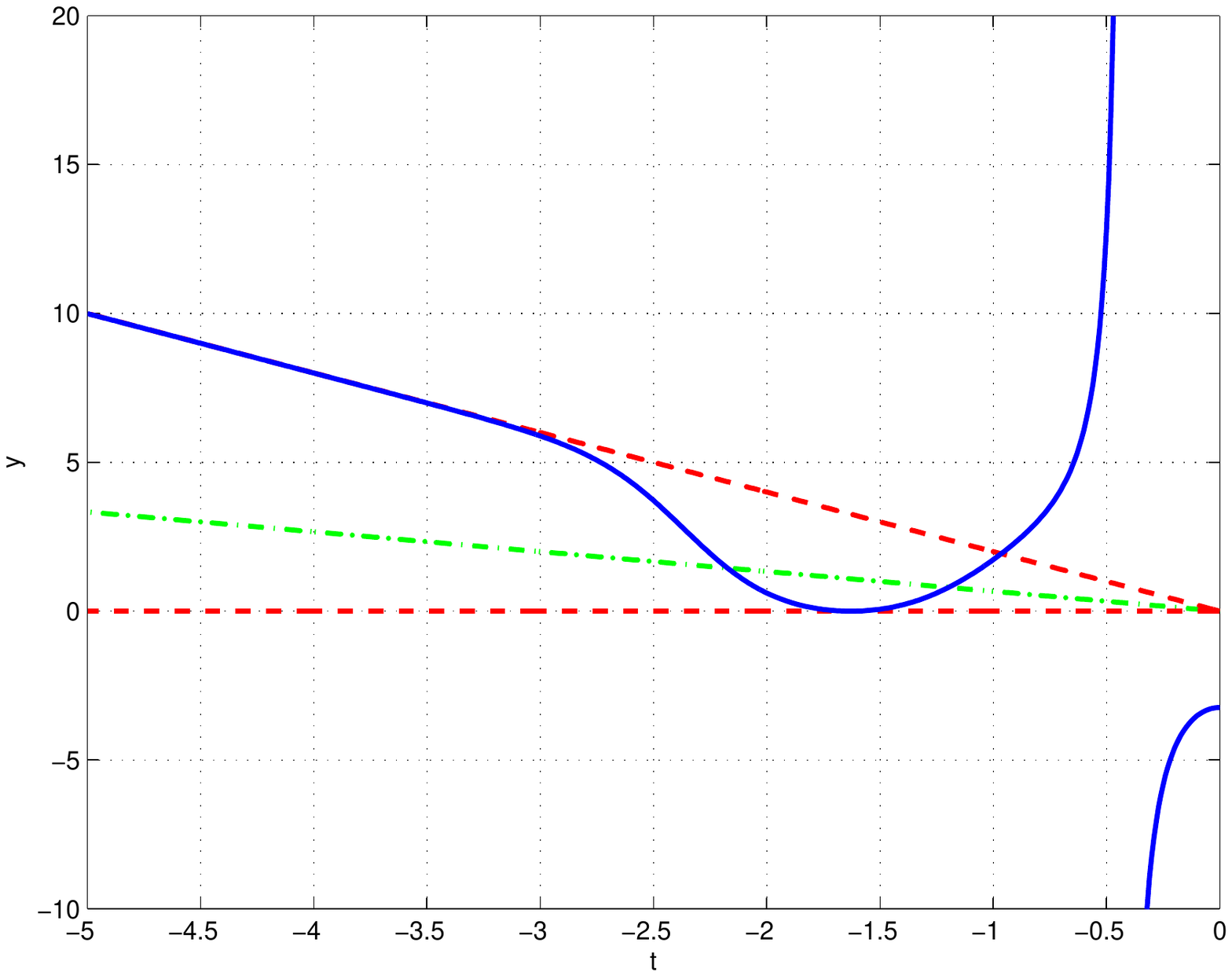}
\hspace{1.5cm}
\end{center}
\null\vspace{-21mm}
\caption{[Color online] First two separatrix solutions (eigenfunctions) of
Painlev\'e IV with fixed initial slope $y'(0)=0$. Left panel: $y(0)=c_1=
-1.98740393$; right panel: $y(0)=c_2=-3.23535569$. The dashed curves are $y=-2t$
and $t=-2t/3$.}
\label{F6}
\end{figure}

\begin{figure}[h!]
\null\vspace{-9mm}
\begin{center}
\includegraphics[trim=21mm 50mm 18mm
50mm,clip=true,scale=0.45]{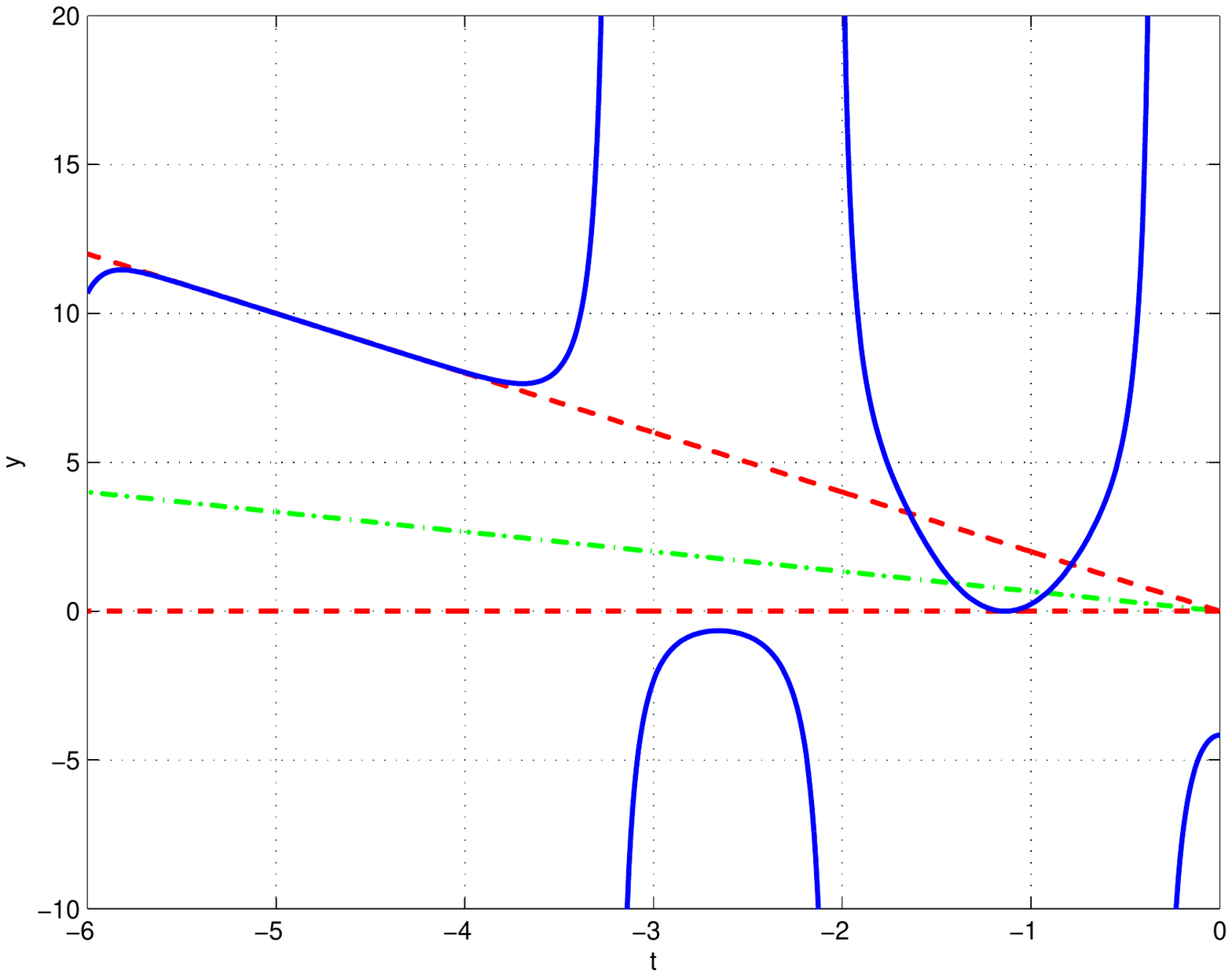}
% trim=left bottom right top
\includegraphics[trim=18mm 50mm 21mm
50mm,clip=true,scale=0.45]{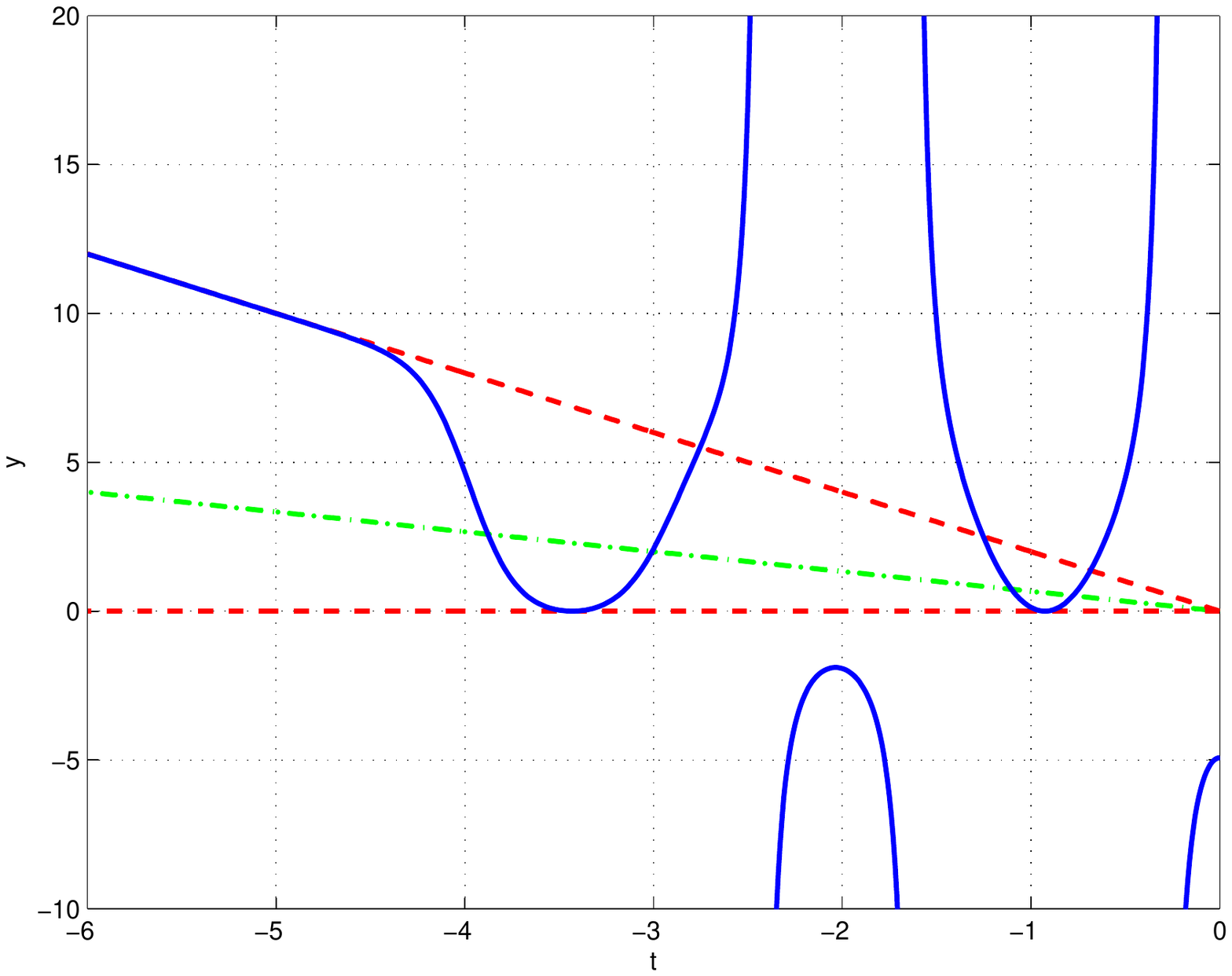}
\hspace{1.5cm}
\end{center}
\null\vspace{-21mm}
\caption{[Color online] Third and fourth eigenfunctions of Painlev\'e IV with
initial slope $y'(0)=0$. Left panel: $y(0)=c_3=-4.1616081$; right panel: $y(0)
=c_4=-4.91908695$.}
\label{F7}
\end{figure}

\begin{figure}[h!]
\null\vspace{-9mm}
\begin{center}
\includegraphics[trim=21mm 50mm 18mm
50mm,clip=true,scale=0.45]{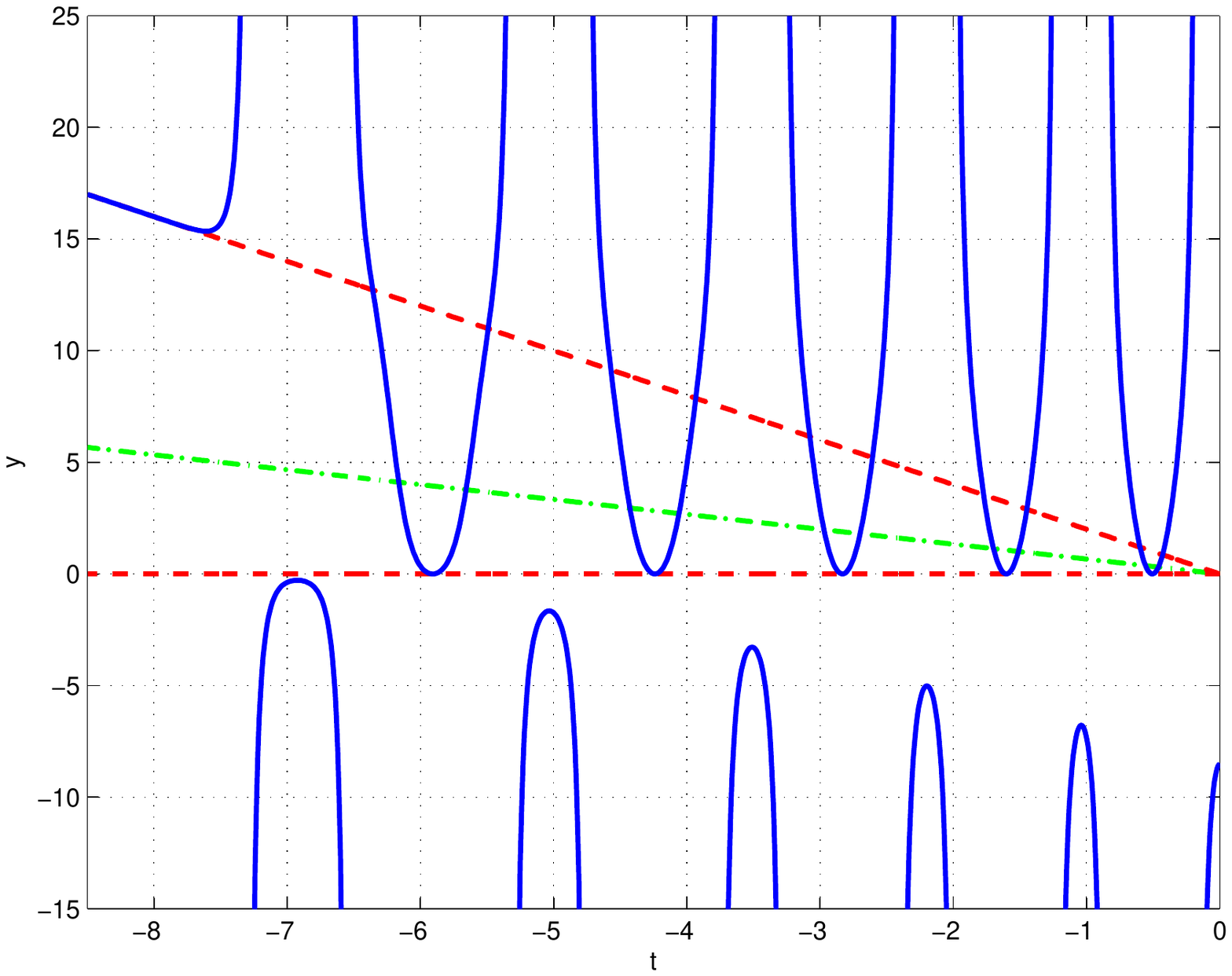}% trim=left bottom right top
\includegraphics[trim=18mm 50mm 21mm
50mm,clip=true,scale=0.45]{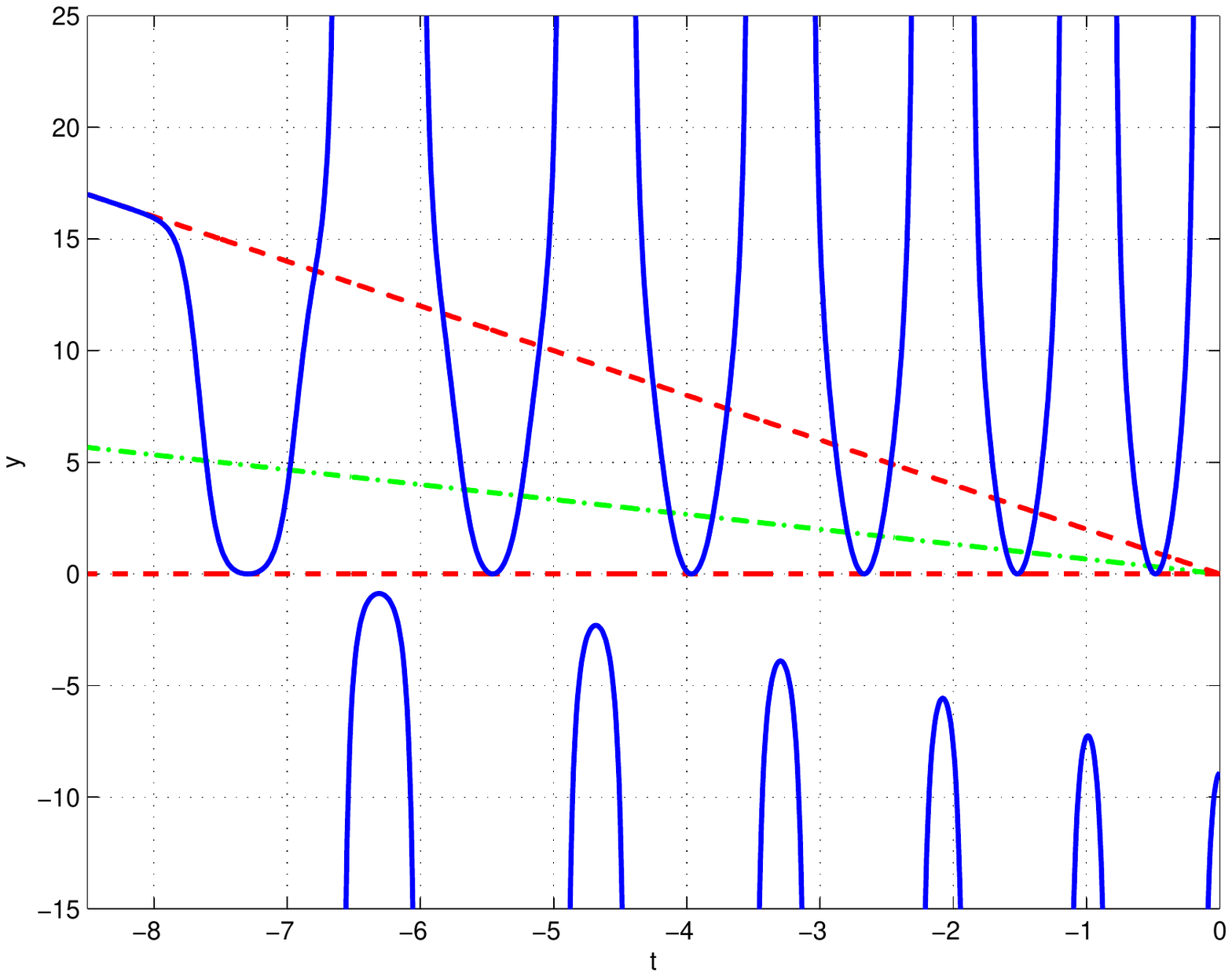}
\hspace{1.5cm}
\end{center}
\null\vspace{-21mm}
\caption{[Color online] Eleventh and twelth eigenfunctions of Painlev\'e IV with
initial slope $y'(0)=0$. Left panel: $y(0)=c_{11}=-8.51211189$; right panel:
$y(0)=c_{12}=-8.90805963$.}
\label{F8}
\end{figure}

If we fix the initial slope at $y'(0)=0$ and allow the initial value $y(0)=c$ to
become increasingly negative, we find a sequence of negative eigenvalues $c_n$
for which the solutions behave like the separatrix (eigenfunction) solutions in
Figs.~\ref{F3}--\ref{F5}. The first two eigenfunctions are plotted in
Fig.~\ref{F6}, the next two in Fig.~\ref{F7}, and the eleventh and twelth in
Fig.~\ref{F8}.

Applying fourth-order Richardson extrapolation to the first 15 eigenvalues, we
find that for large $n$ the sequence of initial-value eigenvalues $c_n$ is
asymptotic to $C_{\rm IV}n^{1/2}$, where
\begin{equation}
C_{\rm IV}=-2.62658{\underbar 7}.
\label{e7}
\end{equation}

\section{Asymptotic determination of $B_{\rm VI}$ and $C_{\rm IV}$} \label{s3}
In this section we present an asymptotic analysis that yields analytic formulas
for $B_{\rm IV}$ and $C_{\rm IV}$ in (\ref{e6}) and (\ref{e7}). To begin, we
rewrite the P-IV equation (\ref{e1}) as
$$2[y(t)]^{3/2}\left[\sqrt{y(t)}\right]''=2t^2[y(t)]^2+4t[y(t)]^3
+\frac{3}{2}[y(t)]^4.$$
This suggests the substitution $u(t)=\sqrt{y(t)}$, which gives the equation
$$u''(t)=t^2u(t)+2t[u(t)]^3+\frac{3}{4}[u(t)]^5.$$
Following Ref.~\cite{R2} we multiply by $u'(t)$ and integrate from $t=0$ to
$t=x$:
\begin{equation}
H\equiv-\half[u'(x)]^2+\eighth[u(x)]^6=-\half[u'(0)]^2+\eighth[u(0)]^6-I(x),
\label{e8}
\end{equation}
where $I(x)=\int_0^x dt\left(t^2u(t)u'(t)+2t[u(t)]^3u'(t)\right)$. The path of
$t$ integration used here is like that used to compute $y(t)$ numerically in
Sec.~\ref{s2}; the path follows a straight line until it approaches a pole, at
which point it makes a semicircular detour in the complex-$t$ plane to avoid the
pole.

\begin{figure}[t!]
\begin{center}
\includegraphics[trim=1mm 5mm 1mm 35mm,clip=true,scale=1.05]{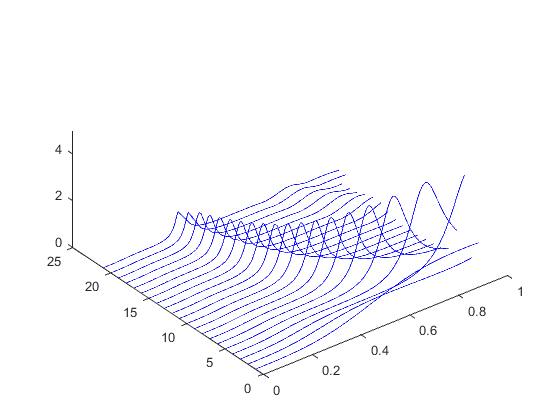}
% trim=left bottom right top
\end{center}
\caption{[Color online] Numerical evidence that $I(x)$ is small for fixed $x$
as $n\to\infty$.}
\label{F9}
\end{figure}

If we evaluate $H(x)$ on the imaginary-$t$ axis we obtain the Hamiltonian
\begin{equation}
\hat H=\half{\hat p}^2+\eighth{\hat x}^6.
\label{e9}
\end{equation}
This Hamiltonian can be interpreted in two possible ways, either as a Hermitian
Hamiltonian for which the eigenfunctions vanish as $x\to\pm\infty$ or as a
$\cPT$-symmetric Hamiltonian for which the eigenfunctions vanish as $|x|\to
\infty$ with ${\rm arg}\,x=-\quarter\pi$ and $-\threequarters\pi$. To see which
quantization scheme is correct we calculate $I(x)$ numerically (see
Fig.~\ref{F9}).

We find numerically that on the lines ${\rm arg}\,x=-\quarter\pi$ and
$-\threequarters\pi$ the function $I(x)$ and becomes small compared with $H$ as
$n\to\infty$ for fixed $x$. Thus, for an eigenfunction of P-IV we can interpret
$H$ as a time-independent quantum-mechanical Hamiltonian. We conclude that the
large-$n$ (semiclassical) behavior of the P-IV eigenvalues can be determined by
solving the {\it linear} quantum-mechanical eigenvalue problem $\hat H\psi=E
\psi$, where $\hat H=\half{\hat p}^2+\frac{1}{8}{\hat x}^6$. The large
eigenvalues of this Hamiltonian can be found by using the complex WKB techniques
discussed in detail in Ref.~\cite{R11}. For the general class of
$\cPT$-symmetric Hamiltonians $\hat H=\half \hat p^2+g\hat x^2\left(i\hat x
\right)^\varepsilon$ $(\varepsilon\geq0)$, the WKB approximation to the $n$th
eigenvalue $(n\gg1)$ is given by
\begin{equation}
E_n\sim\frac{1}{2}(2g)^{2/(4+\varepsilon)}\left[\frac{\Gamma\left(\frac{3}{2}+
\frac{1}{\varepsilon+2}\right)\sqrt{\pi}\,n}{\sin\left(\frac{\pi}{\varepsilon+2}
\right)\Gamma\left(1+\frac{1}{\varepsilon+2}\right)}\right]^{(2\varepsilon+4)/
(\varepsilon+4)}.
\label{e10}
\end{equation}
Thus, for $H$ in (\ref{e9}) we take $g=1/8$ and $\varepsilon=4$ and obtain
the asymptotic behavior
\begin{equation}
E_n\sim\left[\sqrt{\pi}\Gamma\left(\textstyle{\frac{5}{3}}\right)n/
\Gamma\left(\textstyle{\frac{7}{6}}\right)\right]^{3/2}\quad(n\to\infty).
\label{e11}
\end{equation}

Since $\hat H$ in (\ref{e9}) is time independent, we can evaluate $H$ in
(\ref{e8}) for fixed $y(0)$ and large $y'(0)=b_n$ and obtain the result that
\begin{equation}
b_n\sim 4\sqrt{E_n/2}=B_{\rm IV}n^{3/4}\quad(n\to\infty),
\label{e12}
\end{equation}
which verifies (\ref{e5}). We then read off the analytic value of the constant
$B_{\rm IV}$:
\begin{equation}
B_{\rm
IV}=2^{3/2}\left[\sqrt{\pi}\Gamma\left(\textstyle{\frac{5}{3}}\right)/
\Gamma\left(\textstyle{\frac{7}{6}}\right)\right]^{3/4},
\label{e13}
\end{equation}
which agrees with the numerical result in (\ref{e6}). Also, if we take the
initial slope $y'(0)$ to vanish and take the initial condition $y(0)=c_n$ to be
large, we obtain an analytic expression for $C_{\rm IV}$,
\begin{equation}
C_{\rm IV}=-2\left[\sqrt{\pi}\Gamma\left(\textstyle{\frac{5}{3}}\right)/
\Gamma\left(\textstyle{\frac{7}{6}}\right)\right]^{1/2},
\label{e14}
\end{equation}
which agrees with the numerical result in (\ref{e7}).

\section{Concluding remarks} \label{s4}
In this paper we have shown that the fourth Painlev\'e equation P-IV exhibits
instabilities that are associated with separatrix solutions. The initial
conditions that give rise to these separatrix solutions are eigenvalues. We have
calculated the semiclassical (large-eigenvalue) behavior of the eigenvalues in
two ways, first by using numerical techniques and then by using asymptotic
methods to reduce the initial-value problems for the nonlinear P-IV equation
(\ref{e1}) to the linear eigenvalue problem associated with the time-independent
Schr\"odinger equation for the $\cPT$-symmetric $x^6$ potential. The agreement
between these two approaches is exact.

The obvious continuation of this work is to examine the three remaining
Painlev\'e equations, P-III, P-V, and P-VI, to see if there are instabilities,
separatrices, and eigenvalues for these equations as well. It is quite
surprising that P-I, P-II, and P-IV are associated with the $\cPT$-symmetric
$x^2(ix)^\varepsilon$ for the values $\varepsilon=1$, 2, and 4 and it will be
interesting to see if these more complicated Painlev\'e equations have
associated values of $\varepsilon$ as well.

\begin{acknowledgments}
CMB thanks Dr. Marcel Vonk for informative discussions about the properties of
the Painlev\'e transcendents and he thanks the Simons Foundation, the Alexander
von Humboldt Foundation, and the UK Engineering and Physical Sciences Research
Council for financial support.
\end{acknowledgments}


\begin{thebibliography}{100}

\bibitem{R1} C. M. Bender, A. Fring, and J. Komijani, J. Phys. A: Math. Theor.
{\bf 47}, 235204 (2014).
% "Nonlinear eigenvalue problems"

\bibitem{R2} C. M. Bender and J. Komijani, J. Phys. A: Math. Theor. {\bf 48},
475202 (2015).
 % "Painlev\'e Transcendents and PT-Symmetric Hamiltonians" 

\bibitem{R3} J.~Reeger and B.~Fornberg, Stud. App. Math. {\bf 130}, 108 (2012).
% "Painlev\'e IV with both parameters zero: A numerical study" 108-133

\bibitem{R4} J.~Reeger and B.~Fornberg, Physica D {\bf 280-281}, 1 (2014).
% "Painlev\'e IV" General 1-13 <bengt.fornberg@colorado.edu>

\bibitem{R5} D.~J.~Fern\'andez and J.~L.~Gonz\'alez, Ann. Phys. {\bf 359},
213 (2015).
% "Complex oscillations and Painlev\'e IV equation" 213-229

\bibitem{R6} J. Schiff and M. Twiton, J. Phys. A: Math. Theor. {\bf 52},
145201 (2019) and arXiv:1905.12125.
% "A dynamical systems approach to the fourth Painleve equation" -and-
% "Classication of real solutions of the fourth Painleve equation"

\bibitem{R7} C. M. Bender, J. Komijani, and Q-h. Wang, in {\it Resurgence,
Physics and Numbers}, ed. by F. Fauvet, D. Manchon, S. Marmi, and D. Sauzin,
CRM Series, Ennio De Georgi {\bf 20}, 67-89 (2017).

\bibitem{R8} C. M. Bender, J. Komijani, and Q-h. Wang, J. Phys. A: Math.
Theor. {\bf 52}, 315202 (2019).
% "Nonlinear eigenvalue problems for generalized Painlev\'e equations"

\bibitem{R9} O. S. Kerr, J. Phys. A: Math. Theor. {\bf 47}, 368001 (2014).
% "Comment on `Nonlinear eigenvalue problems'"

\bibitem{R10} C.~M.~Bender and S.~A.~Orszag, {\it Advanced Mathematical
Methods for Scientists and Engineers} (McGraw Hill, New York, 1978).

\bibitem{R11} C.~M.~Bender and S.~Boettcher, Phys. Rev. Lett. {\bf 80}, 5243
(1988).
% "Real Spectra in non-Hermitian Hamiltonians having PT symmetry" 5243-5246

\end{thebibliography}
\end{document}